\documentclass[12pt]{article}
\usepackage[utf8]{inputenc}
\usepackage[margin=1in]{geometry}
\usepackage{setspace}
\usepackage{authblk}
\usepackage[hidelinks]{hyperref}
\usepackage{graphicx}

\usepackage[american]{babel}

\usepackage[backend=biber,style=apa, maxnames = 6, maxbibnames = 999]{biblatex}
\raggedright

\DefineBibliographyExtras{english}{}

\begin{document}

\title{Towards Immersive Generosity: The Need for a Novel Framework to Explore Large Audiovisual Archives through Embodied Experiences in Immersive Environments}
\author[1,*]{Giacomo Alliata}
\author[1]{Sarah Kenderdine}
\author[1,2]{Lily Hibberd}
\author[3]{Ingrid Mason}
\affil[1]{Laboratory of Experimental Museology, EPFL}
\affil[2]{UNSW Sydney}
\affil[3]{Australian National University}
\affil[*]{Corresponding author: giacomo.alliata@epfl.ch - Rue des Jordils 41, St-Sulpice, Switzerland}
\setcounter{Maxaffil}{0}
\renewcommand\Affilfont{\itshape\small}
\date{}

\maketitle

\renewenvironment{abstract}
 {\begin{center}
  \bfseries \abstractname\vspace{-.5em}\vspace{0pt}
  \end{center}
  \list{}{%
    \setlength{\leftmargin}{0cm}
    \setlength{\rightmargin}{\leftmargin}%
  }%
  \item\relax}
 {\endlist}

\begin{abstract}
  This article proposes an innovative framework to explore large audiovisual archives using Immersive Environments to place users inside a dataset and create an embodied experience. It starts by outlining the need for such a novel interface to meet the needs of archival scholars and the GLAM sector, and discusses issues in the current modes of access, mostly restrained to traditional information retrieval systems based on metadata. The paper presents the concept of ``generous interfaces" as a preliminary approach to address these issues, and argues some of the key reasons why employing Immersive Visual Storytelling might benefit such frameworks. The theory of embodiment is leveraged to justify this claim, showing how a more embodied understanding of a collection can result in a stronger engagement for the public. By placing users as actors in the experience rather than mere spectators, the emergence of narrative is driven by their interactions, with benefits in terms of engagement with the public and understanding of the cultural component. The framework we propose is applied to two existing installations to analyze them in-depth and critique them, highlighting the key directions to pursue for further development.
\end{abstract}

\section{Introduction}

At the beginning of his book \emph{Zen and the Art of Motorcycle Maintenance}, Robert \textcite{pirsig} presents the difference felt while travelling by car compared to on a motorbike. In the first case, travelers are watching the environment through their car windows, while in the second, they are ``feeling", ``living" that same environment. The humidity in the air, the wind on their faces, the vivid sound of the motor, the unconstrained visual field, all contribute to create a more embodied and immersive experience for the motorcyclists. A comparable experience is described by \textcite{slater_framework_1997}, in which one of the authors, who, after several years of looking through the windows of a computer room in passing, is one day obliged to enter and is confronted with the shock of suddenly going ``inside" what had only ever been seen from the ``outside". This short story illustrates what Immersive Environments (IEs) offer to their users: the possibility to be ``inside" and therefore to feel as if you were ``there", a more fully embodied experience, (more or less) completely surrounded by a virtual world. Often described as the sense of ``being there", this psychological state is sometimes also referred to as ``presence" and is a concept that is one of the central features of IEs \parencite{steuer_1992}. ``Immersion" is arguably the second central feature of IEs: the extent to which the system can deliver ``an inclusive, extensive, surrounding and vivid illusion of reality to the senses of a human participant" \parencite[][p. 1]{slater_framework_1997}. Since the display size remains a critical limitation in the presentation and understanding of data visualizations \parencite{kasik2009data}, many large display systems (more or less immersive) have been developed in the last decades, such as wide gigapixel screens, dome-like structures, CAVEs and half-CAVEs or panoramic screens (see \textcite{huhtamo2013illusions} for an historical perspective on IEs). 

Despite this progress, the exploration of cultural collections in IEs faces a range of limitations and dilemmas, for which a new framework is required. This article proposes that such a new paradigm is possible and that, within the project \emph{Narratives from the Long Tail: Transforming Access to Audiovisual Archives}, the foundations for this innovative framework are being laid and will be applied to important audiovisual collections to create new ways to explore them\footnote{Details on the archives explored can be found at \url{https://www.futurecinema.live/archives-and-collections/}.}. \emph{Narratives}'s goal is to produce a ``groundbreaking visualization framework for interactively (re)discovering hundreds of thousands of hours of audiovisual materials" \parencite{kenderdine2021computational}. The remainder of this introduction offers a summary of the key concepts and technologies to be discussed.

As the foremost mnemonic records of the 21st century, audiovisual recordings are omnipresent in our daily lives. From the second half of the 20th century, starting with the introduction of television, to today's online sharing platforms, such as Youtube or TikTok, audiovisual recordings play an important role in the way we document, disseminate and preserve knowledge and culture. Broadcasting institutions are digitizing their collections, with examples such as the Radio Télévision Suisse (RTS) with 200,000 hours of footage \parencite{rtsarchive}, or the British Broadcasting Corporation (BBC) with more than a million recorded hours \parencite{wright_bbc}. Furthermore, cultural video collections are useful to preserve Intangible Cultural Heritage (ICH), which has been defined as ``the culture that people practise as part of their daily lives" \parencite{kurin} and ``all [the] immaterial manifestations of culture [that] represent the variety of living heritage of humanity as well as the most important vehicle of cultural diversity" \parencite{lenzerini}. One can think of events like rituals, dance performances or festivals that audiovisual collections, due to their temporal value as well as the combination of video and audio channels, are more apt to represent than images or textual elements ever could. Audiovisual archives by themselves, however, can only offer a mostly linear, single-channel, non-interactive narrative. They need to be augmented by some sort of interface to be explored in a meaningful way.

In the field of cultural heritage, the need for innovative ways to present such collections to the public is evident. A 2014 survey conducted on 1200 cultural institutions in Europe established that around 80\% have digitized their collections \parencite{survey_digital_collections}, and this number has certainly increased since then. The Europeana platform\footnote{\url{https://www.europeana.eu/en}} is, for instance, a key example, with more than 30 million items accessible online through a category-based interface. The majority of these institutions provide online access to (at least part of) their collections, and even though innovative forms of access can be noted \parencite{windhager_visualization_2019}, one can argue that these largely remain constrained to the web. Web-based systems, such as Europeana's, are obviously powerful tools to enhance access to these collections and democratisation of culture, however they lack the immersion that IEs installations can offer. These are especially relevant when showcasing collections in public spaces such as museums or cultural institutions venues. An innovative framework to explore large digital collections is therefore necessary, in particular, to present audiovisual archives to a larger public.

Within IE theory, one key aspect to consider is the way narrative can emerge in an embodied experience, what can be referred to as Immersive Visual Storytelling (IVS). This term comprises the ensemble of processes and approaches that are employed to generate narratives in immersive experiences (see \textcite{llobera2013telling,layng2019cave}, for some examples). It is on this basis that we claim that, through this emergence of narrative, users can better understand and explore large cultural collections, by being placed ``inside" the dataset \parencite{shen-2019} and being offered levels of interactivity to freely browse the collection, as well as making them actors in the experience rather than mere spectators, resulting in a more ``embodied understanding" of the content displayed \parencite{johnson2015embodied}. 

The next part of this paper justifies the need for a new framework to explore large audiovisual collections (and cultural collections, in general), drawing on Whitelaw's ``generous interface" theory \parencite{whitelaw_generous_2015} and outlining the benefits immersion can bring to the exploration of such archives. The relevant theories of embodiment will then be presented and discussed in relations to IEs and IVS with specific examples, in order to understand how ``embodied understanding" works and therefore how it can be leveraged to enhance the experience. Finally, two previous works will be analyzed in-depth and critiqued in light of the proposed framework: \emph{T\_Visionarium II}, developed at UNSW's iCinema Research Centre in 2006, and \emph{Jazz Luminaries}, created in 2019 at EPFL's Laboratory of Experimental Museology. In each case, users are immersed within large audiovisual datasets, and this results in an embodied understanding of the archives, from which narrative can emerge.

\section{The Need for a New Framework: Towards Immersive Generosity}

From the perspective of scholarly archival community, it is clear that large audiovisual archives are currently lacking proper frameworks to explore them. These collections remain mostly inaccessible because cultural institutions are constrained to screening sessions where only a handful of videos can be shown at one time, without revealing the dataset in its entirety. The Digital Humanities and GLAM sector (galleries, libraries, archives, and museums) have also called for innovative forms of engagement through compelling frameworks to explore this kind of collections \parencite{fossati-2012}. 

For one, the classical information retrieval system restrains users to classifications and tagging prior interpretations provided by curators and data managers. Furthermore, the metadata compiled prevents users from accessing visual features because images are by nature harder to verbally capture \parencite{masson-2020}. A ``simple search" interface is thus insufficient to provide compelling public engagement with the diverse array of media formats held in cultural institutions \parencite{rogers2014doesn}. The framework proposed in this article intends to solve this problem, as it provides users new ways to access large collections, thus dispensing with the need to rely on metadata and traditional cataloguing forms of description.

\begin{figure}[!h]
 \centering
 \includegraphics[width = 0.9\textwidth]{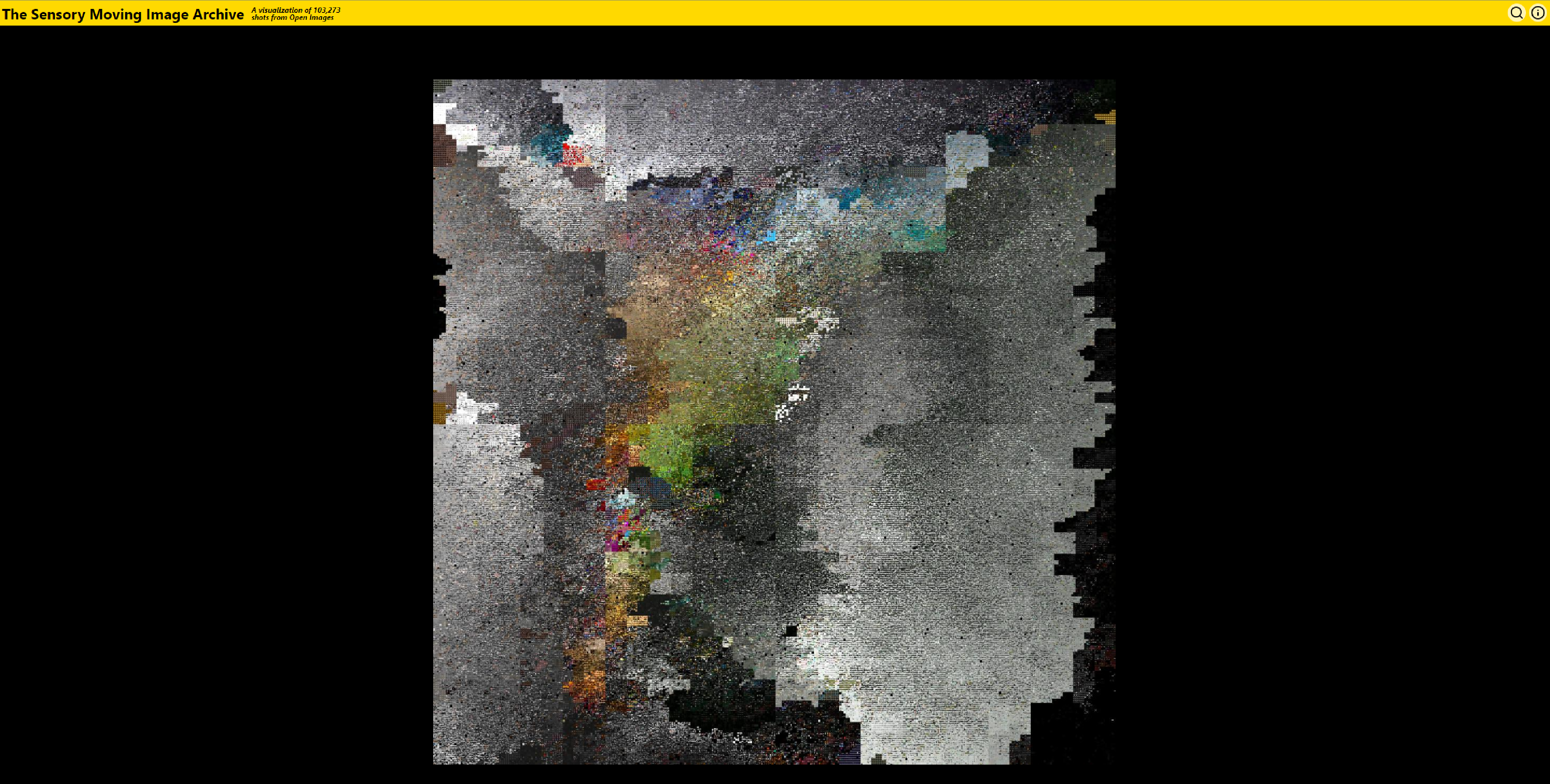}
 \caption{Main view of the SEMIA application. Shots are spatially distributed by color using a t-SNE algorithm of the similarity between all shots (credits: Bert Spaan).}
 \label{fig:semia}
\end{figure}

A catalogue-like database description of an audiovisual collection cannot fully describe its content because discrete categories do not encompass the visual and temporal complexity of videos. More continuous analytics can in theory characterize visual features of a video but in practice these seem to be harder to capture as they require some sort of interpretation. One could imagine a scale from 0 to 1 measuring the visual complexity of a shot, for instance. However, a catalogue of entries between 0 and 1 would be much harder to interpret than actually watching the related videos and visually compare them, as in the SEMIA\footnote{\url{https://bertspaan.nl/semia/\#/}} project. This recent initiative undertaken at the University of Amsterdam is a key example of modern web-based approaches to the exploration of audiovisual archives \parencite{masson-2020}. Figure \ref{fig:semia} presents the main view of the application, in which 103,273 shots from 6,969 videos are spatially distributed in 2-D based on similarities on visual features: color, shape, movement or visual complexity (as in the example above). With this interface, one can easily appreciate comparable shots and grasp the full collection at a glance, an impossible task with just a list of catalogue entries.

Furthermore, traditional access to cultural collections shapes the memories we create of the past. These memories ``are not inherent in the archival stock, but are created in the context of reception, through processes of remediation and recontextualisation" \parencite{brunow2017curating}. This relates to the idea that traditional metadata rely on prior assumptions of their authors, as well as on the initial goal of these descriptions. For an archival entity, a catalogue is first and foremost a way to document their collections, usually connected to a database system with an information retrieval tool to (at least internally) find the relevant items. This purpose is quite different than browsing through a collection without any prior knowledge (a goal that seems to be relevant for casual users, \parencite{lopatovska2013exploring} and therefore solely relying on these metadata might not be enough. Browsing is indeed ``a rich and fundamental human information behaviour" \parencite{chang1993browsing}, an iterative process based on scanning \parencite{rice2001accessing} or glimpsing \parencite{bates2007browsing} a collection of items. These processes also depend on the questions one might ask, and it has been shown that revealing a collection in its entirety through the use of spatial distributions for instance can prompt innovative questions \parencite{olesen2016data}. People ``browse with or without a goal in mind, and goals may change as the process unfolds" \parencite{whitelaw_generous_2015}.

Visualization is therefore at the center of an innovative framework to explore large audiovisual collections. Visualization, as ``a medium for communication (or persuasion, or engagement)" and a tool for ``understanding (or problem solving, planning, orienting)" \parencite[][pp. 1-2]{scagnetti2011visual}, can reveal patterns, structures, relationships in the data and prompt new enquiries. Much like Moretti's ``distant reading" approach to literature, which can disclose hidden meanings in the text \parencite{moretti2013distant}, visualization can expose new knowledge. For design and humanities scholar Johanna Drucker, visualization ``produces" knowledge through ``graphical forms expressing interpretation", and that because of the ``fundamentally interpreted condition on which data is constructed" visualizations are a feature of both ``knowledge production and [its] presentation" \parencite{drucker2014graphesis}.

Shneiderman's ``visual information seeking" approach entails a taxonomy of tasks users might want to perform while exploring a collection: overview, zoom, filter, details-on-demand, relate, history, and extract \parencite{shneiderman_eyes_2003}. This interaction paradigm requires surrogates in the form of previews for single items and overviews for groups of items \parencite{greene_previews_2000}, to represent the collection objects while exploring it along its latent dimensions. According to \textcite{drucker_performative_2013}, however, Shneiderman's information visualization ``mantra" is ``pragmatic, but highly mechanistic" and supposes users with clear goals in mind, something that might not actually hold true for casual visitors in a museum space. This ubiquitous task-based approach, widespread in the fields of human-computer interaction and information retrieval, does not meet the criteria of a more humanistic approach. It is furthermore essential to recognize the inherent reward component and creativity aspect in the action of browsing a collection \parencite{rice2001accessing}. While browsing, a user is the sole director of the experience (although they might be more or less consciously guided towards a certain path) and is therefore creating their own narrative. It is a very different thing to go through a curated list of items and discover the same items while autonomously exploring the collection. In the latter case, serendipity, ``the fact of finding interesting or valuable things by chance" \parencite{cambridge_dict}, plays a major role. The feeling of finding new items by chance, of ``serendipitous" discoveries \parencite{toms2000serendipitous}, entails a procedural emergence of narrative, driven by the user-agent of the installation. 

All these concepts are summarized by Whitelaw's innovative notion of ``generous interfaces" \parencite{whitelaw_generous_2015}. As he argues, searching requires ``rich, browsable interfaces that reveal the scale and complexity of digital heritage collections". It should be a ``humanistic model of interface and interaction that emphasises exploration and interpretation over task and information retrieval". The modes of visual storytelling offered by these generous interfaces are simultaneously ``horizontal" (through the browsing features) and ``vertical" (through the details-on-demand functionalities), more or less completely driven by the user. Whitelaw's interfaces empower users to generate, to craft new knowledge, through the narrative they are creating. Psychologist Jerome Bruner discusses the importance of narrative for its fundamental role in creating and interpreting human culture \parencite{acts_of_meanings}. He states that human beings are natural storytellers: they make sense of the world and themselves through narrative. From the time they are very young, children learn that the way to integrate their own desires with their family’s norms and rules is to construct a story about their actions. This push to construct narrative, Bruner maintains, shapes how children acquire language, and the habit persists into adulthood as a primary instrument for making meaning. These storytelling skills insure our place within human society. This point is sustained by constructionist theories \parencite{dewey-1966, piaget-1973}, according to which individuals do not learn by passively perceiving content but rather by actively crafting, manipulating and therefore creating new knowledge.

Through their rich browsing features and interactivity, generous interfaces evidently offer ideal modes of access to large digital collections, which are far more interesting than traditional information retrieval systems. These interfaces are however mainly web-based and therefore restrained to single users in front of small and flat screens. One could argue the immersive component of IVS is completely lacking here, as well as the multi-user aspect. The framework we suggest would solve these issues through the use of Immersive Environments, relying on the concept of an \textit{immersive generosity}. 

By transposing the generous interface concept to IEs, one must however consider how it will affect the narrative. Indeed, a story is closely correlated with the medium used to convey it. This is perfectly illustrated with cinematic adaptations of books: the overall story being told is perhaps the same, but the way it is told, its content and its intensity can greatly differ. Applying this idea to IVS, it is clear that the active role a user has in an immersive and embodied application vastly influences the way narrative can emerge, requiring a distinction between purely authorial storytelling and interactive approaches. \textcite{aylett_towards_2003} propose a useful model for this with four dimensions to characterize the narrative component of different mediums: ``Contingency" (the contingency of time and space of the story being told with respect to the real time and space of the user); ``Presence" (how much the user feels present in the story); ``Interactivity" (the degree of controls they have on the narrative) and ``Narrative Representation" (the form narrative adopts, be it through mental models for literature for instance or purely visual and aural for cinema). They argue that, when compared to the most common cases of narrative mediums, namely literature, cinema and theatre, virtual reality offers the greater contingency in time and space, the strongest feeling of ``being there" and the highest degree of interactivity. 

Combined, these considerations imply that IVS is a form of narrative that moves beyond the Platonic concepts of ``diegesis" and ``mimesis" (based on an authorial view of storytelling) to the idea of ``experiencing" and ``creating" a story. 
When focusing on multi-user experiences, typically offered by large interactive systems, the concept of a user-led narrative has even greater implications for the way the other users in the interactive space experience the story being created. In Geert Mul's work, users can simultaneously be seen as ``highly productive, in that the appearance of the works changes based on their input" and ``merely one in a much larger series of variables that determine the outcome of the calculation" \parencite{mul-2018}. Although these views might seem at odds, they both imply that, when IVS is user-led, it requires an external public (other users) to interpret it and appreciate it fully, through a ``third-person's perspective". One must also remember that museums are historically social venues, and the relationships between visitors are intrinsic to the experience of exploring their collections, meaning that these principles must be applied to to the exploration of large datasets in multi-user immersive installations. 

In conclusion, Whitelaw's ``generous interfaces" have been highly influential as a first attempt to solve the issues outlined by archival and digital humanities scholars in the access to large audiovisual collections. Nonetheless, we argue that these interfaces would benefit from a further component of immersion, through the use of IEs and thus IVS, to create a truly embodied exploration of a cultural archive. To conceptually frame this idea, philosopher Mark Johnson's theory of the body is leveraged in the next section to better frame how narrative can emerge from such installations through ``embodied understanding" \parencite{johnson2015embodied}.

\section{Embodied Understanding in Immersive Environments}

For \textcite{johnson2015embodied}, the importance of ``embodied understanding", based on the 20th century findings of cognitive science, challenges centuries during which the body was considered less important than the mind. He provides the counter-argument that ``understanding is profoundly embodied, insofar as our conceptualization and reasoning recruit sensory, motor, and affective patterns and processes to structure our understanding of, and engagement with, our world". It is therefore clear that IVS, through its embodied approach, can indeed be beneficial to cultural institutions aiming to give the wider public meaningful access to their collections.

To put these ideas in practice in IEs-based installations, one must however first appreciate how our understanding of the world is embodied. According to the field of embodied cognition, organism-environment interactions are the sources of all our human perception and understanding of the world \parencite{dewey1981}. To really understand something, we must first experience it, a complex process based on Damasio's ``homeostatis" balanced state between organism and environment that comprises both how the organism is feeling and acting as well as how the environment is structured \parencite{damasio2010self}. This equilibrium is dynamic, because it evolves as the organism and the environment evolve, and is also related to the quality, the value of the experience and overall our well-being. This is why emotions are such an important element of our ``embodied understanding" of the world. Neuroscientists have further shown how organisms, through the detection of ``emotionally competent stimulus", move their body-states to favorable positions for their survival and well-being \parencite{damasio2003looking}. Emotions are therefore at the core of understanding, something that seems quite obvious in storytelling, as any story plays with our emotions to convey its narrative.

Going one step further, \textcite{johnson_what_2008} talks of the ``bodily sources of meaning". Humans have indeed always used their bodies to express themselves, from spontaneous gestures while talking to more elaborate performances such as dance or ritual practices \parencite{johnson2007meaning, ziemke2007body}. It is thus necessary to define this ``body" of ours, and relate its dimensions to IVS in the frame of the exploration of large cultural collections. Drawing on Merleau-Pointy's \emph{Phenomelogy of Perception} and John Dewey’s somatic naturalism, Johnson explains how our bodies are not just ``objects interacting with other objects" but are ``lived", ``phenomenal bod[ies]" \parencite{merleau2013phenomenology}, and require at least five intertwined dimensions to be fully comprehended: the biological, ecological, phenomenological, social and cultural body.

Going back to Damasio's homeostatis state, our bodies are first and foremost ``flesh-and-blood", ``functioning biological organism[s] that can perceive, move within, respond to, and transform [their] environments" \parencite{dewey1981}. Our ``biological bodies" are in a continuous exchange with their environments, continuously evolving the aforementioned equilibrium, and are the locus of feelings and emotions that push us towards our physical and social well-being \parencite{damasio2003looking}. When being confronted with an immersive narrative experience, users are therefore first and foremost a biological body, with all their individualities and body specificities. In her interactive and immersive experience \emph{Osmose}\footnote{\url{http://www.immersence.com/osmose/}} (1995), media artist Char Davies empowers visitors to automatically drive the narrative through their breathing and balance, two fundamentals and unconscious human activities. These are part of the preconscious activities \textcite{gallagher2005body} identifies as the ``body schema" that govern our interactions with the environment.

Once users start to actively engage and interact with an installation, the second dimension of the body emerges: the ``phenomenological body". Our ``tactile-kinaesthetic body" \parencite{sheets-1999} depends on proprioception (our feeling of our bodily posture and orientation), our kinaesthetic sensations of bodily movement and our awareness of our internal body states through our emotions and feelings \parencite{damasio2003looking}. In contrast to the body schema, \textcite{gallagher2005body} associates our more conscious activities to the ``body image": activities that comprehend the affordances of the system to explore the collection, Drucker's ``conventions of the diagrammatic knowledge form" \parencite{drucker2014graphesis}. Furthermore, this phenomenological paradigm can result in the enhancement of a cognitive operation through a shift in the nature of the task itself, where abstract operations (such as finding all the items that correspond to a certain query) can be mapped to more natural actions, the so-called ``tangialities" \parencite{milekic2002towards}. 

\begin{figure}[!h]
 \centering
 \includegraphics[width = 0.7\textwidth]{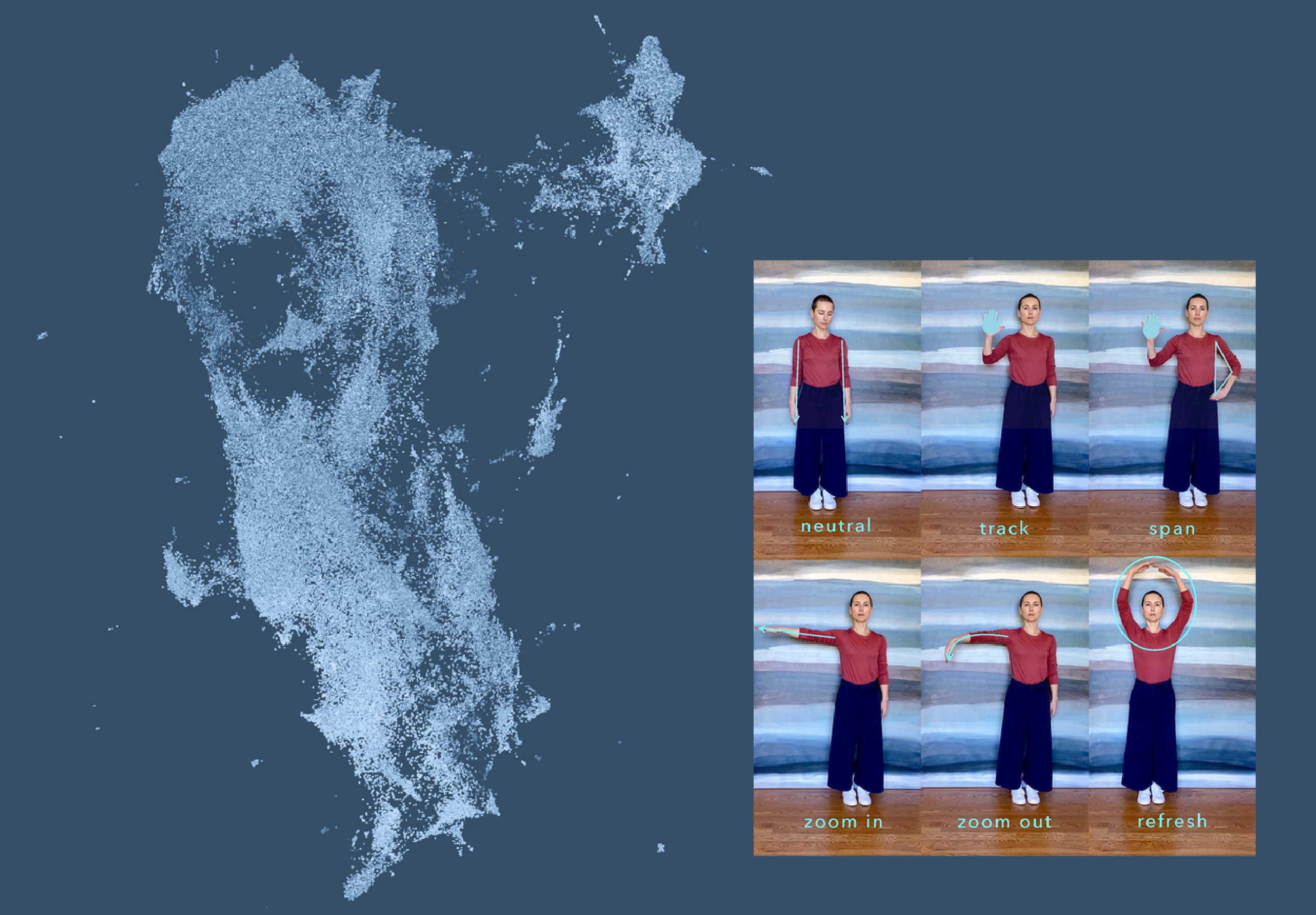}
 \caption{\emph{Surprise Machines} media installation with the museum's collection spatially distributed on the left and the gesture vocabulary on the right (credits: metaLAB (at) Harvard).}
 \label{fig:surprise_machines}
\end{figure}

Dario Rodighiero and his colleagues at metaLab at Harvard have pushed this dimension to the extreme of posing the entire body as the ``interaction device", a sort of ``choreographic interface". In \emph{Surprise Machines}\footnote{\url{https://dariorodighiero.com/Surprise-Machines-for-Harvard-Art-Museums}}, visitors explore Harvard Art Museum collections through the use of precise and choreographed gestures \parencite{surprise_machines}, each one mapped to a specific task that remind us of Shneiderman's visualization mantra: ``overview first, zoom and filter, then details-on-demand" \parencite{shneiderman_eyes_2003}. Figure \ref{fig:surprise_machines} shows the digital collection spatially distributed according to visual similarity of the different items and the gesture vocabulary defined to explore this latent space. 

In this installation, the digital collection is shaped to create a virtual environment that users can freely explore through their bodies. This intrinsic relationship to the environment is elucidated by a third dimension: the ``ecological body", which can be defined as the continuous process between our bodies and the environment we evolve within \parencite{dewey1981, merleau2013phenomenology}. The twofold phenomena of ``embodied understanding’ and the ``ecological body" is pushed even further in IEs such as the EPFL Laboratory of Experimental Museology's Panorama+ or its predecessor, the UNSW's iCinema Research Centre's landmark system Advanced Interaction and Visualization Environment (AVIE). The Panorama+ is a 360-degree stereoscopic, interactive environment of ten meters diameter and four meters high, with five projectors and surround sound audio system, controlled by a cluster of six computers (see \textcite{mcginity_avie_2007} for a more in-depth technical description of the AVIE). Its omnidirectional nature recreates a fully immersive data space that allows for both allocentric (relationships between objects) and egocentric (personal relationships to objects) cognition and spatial perspectives simultaneously \parencite{blesser2009spaces}, cited in \parencite{kenderdine_embodiment_2015}. Visitors can indeed physically represent themselves with respect to digital objects in the virtual landscape and appreciate relationships between these digital items, blurring the line between what is real and what is virtual. This phenomenon increases their sense of presence and thus enhancing the narrative component of the exploratory experience. The digital collection is no longer an ascetic database or list of catalogue entries but a fully-fledged environment that users can explore on their own terms and to which they can individually or collectively relate with other visitors, in the case of multi-user systems.

It is in these multi-user experiences that the fourth dimension of the body Johnson outlines, the ``social body", is revealed. Indeed, our environment is not just physical or biological but also composed of human relationships and interactions with our peers. In the field of developmental psychology, the effect (and importance) of other people during our childhood is well-known \parencite{stern2018interpersonal}, and this continues as adults, through interactions with our colleagues and friends. This is especially true in social spaces such as museums, where users are generally not alone but rather continuously confronted with the presence of others, usually strangers, who become the ``spectators" in the trichotomy ``system-user-spectators" \parencite{dourish2001seeking}. Embodiment, on this basis, can thus be argued to have a ``participatory" status where the user driving the experience becomes the author of narrative for a larger public. Here again, the ``third-person's perspective" Geert Mul mentioned in his work, as well as the dual view of the role of users in generative interactive pieces are crucial \parencite{mul-2018}. From the perspective of performance studies theory, ``it is the ways in which the user perceives and experiences the act of interacting with the system under the potential scrutiny of spectators that greatly influences the interaction as a whole ... it is precisely this awareness of the (potentiality of a) spectator that transforms the user into a performer" \parencite{dalsgaard2008performing}. Similarly, the latest theoretical frameworks on creativity highlight the importance of the ``public" in the creative act \parencite{sternberg-2022}: the interaction with a given system in this case. This creative act of interacting with an installation also has an important effect on the ``embodied understanding" of the collection, as put forward by constructionist theories on learning \parencite{dewey-1966, piaget-1973}. Individuals actively engaged with the knowledge they are being presented will learn more than those passively witnessing it. Bloom's famous taxonomy of educational objectives supports this claim, since its main categories include Application, Analysis, Synthesis and Evaluation (activities that require the manipulation and creation of knowledge), in addition to Knowledge and Comprehension \parencite{bloom1956taxonomy}. One of the more modern revisions of this taxonomy puts emphasis on the notion of creating new knowledge by splitting the original classification into two dimensions: the first on the actual knowledge being addressed, and the second on the cognitive processes applied to this knowledge \parencite{krathwohl2002revision}. Furthermore, empirical evidence suggests the hierarchical characteristic of these taxonomies, placing the category of Create at the top \parencite{anderson2001taxonomy}.

The fifth and last dimension of the body Johnson describes is the cultural one. He argues that various cultural aspects contribute to the shaping of our bodies and the way we see and relate to them. This explains why gestures and postures vary across the world, as well as our attitudes towards our environment. Cultures are enacted through rituals, practices, customs performed by humans as inherently embodied beings \parencite{johnson_what_2008}. Therefore, in IVS, the emergent narrative will be enacted by the embodied users, attended by their prior knowledge and specific cultures. The interpretative action of the visitors mentioned before depends on the individual preconceptions people bring to the experience, and consequently the emergent narrative that results from their interactions can greatly differ. The dialogue between system and users is driven by these interpretations, users with specific backgrounds will draw connections that might appear rather strange to other visitors but nonetheless ``make sense" in their specific dialogue, in their specific narrative. This fundamentally individual interpretative action reverts back to the social body and Mul's ``third-person's perspective", where the individual differences between users spark seemingly infinite combinations and unfolding of different narratives. Viewers' own cultures and previous knowledge become new variables in the process of generating these narratives, both for the user actually driving the experience and for the public interpreting it. Furthermore, the cultural aspect is particularly relevant when exploring audiovisual archives, because the ``immaterial manifestations of culture" (what we refer to as Intangible Cultural Heritage) can be captured and documented through videos, so that exploring such a collection amounts to exploring culture itself (or at least an aspect or portion of it). The power of IEs to ``plac[e] users inside the dataset" \parencite{shen-2019} can thus immerse them in a cultural setting, with the various benefits for their experience, both for their engagement and for their learning. This full immersion in a collection creates an ``embodied theatre of participation" that ``permits an unprecedented level of viewer co-presence in a narrative-discovery of a cultural landscape", facilitating ``dynamic inter-actor participation and cultural learning" \parencite{kenderdine_somatic_2007}. 

The need for an innovative framework having been outlined and said framework being situated within embodiment theory, this thinking will now be applied to analyze and critique two previously-built installations, to highlight the particularities and advantages of our proposition.

\section{Critique of Interactive Installations in Light of the Proposed Framework: \emph{T\_Visionarium II} and \emph{Jazz Luminaries}}

The need for an immersive generosity having been justified and the concept of embodied understanding explained through Johnson's theory of the body, we now illustrate how this innovative framework we propose can be applied to previously built installations created to explore large video collections. These use cases will first be presented and then critiqued in light of the proposed framework, in an attempt to formalize it and draw conclusions on what the next stage of immersive interfaces for exploring large video collections should aim for. The two installations discussed here are \emph{T\_Visionarium II} and \emph{Jazz Luminaries}.

\begin{figure}[!h]
 \centering
 \includegraphics[width = 0.7\textwidth]{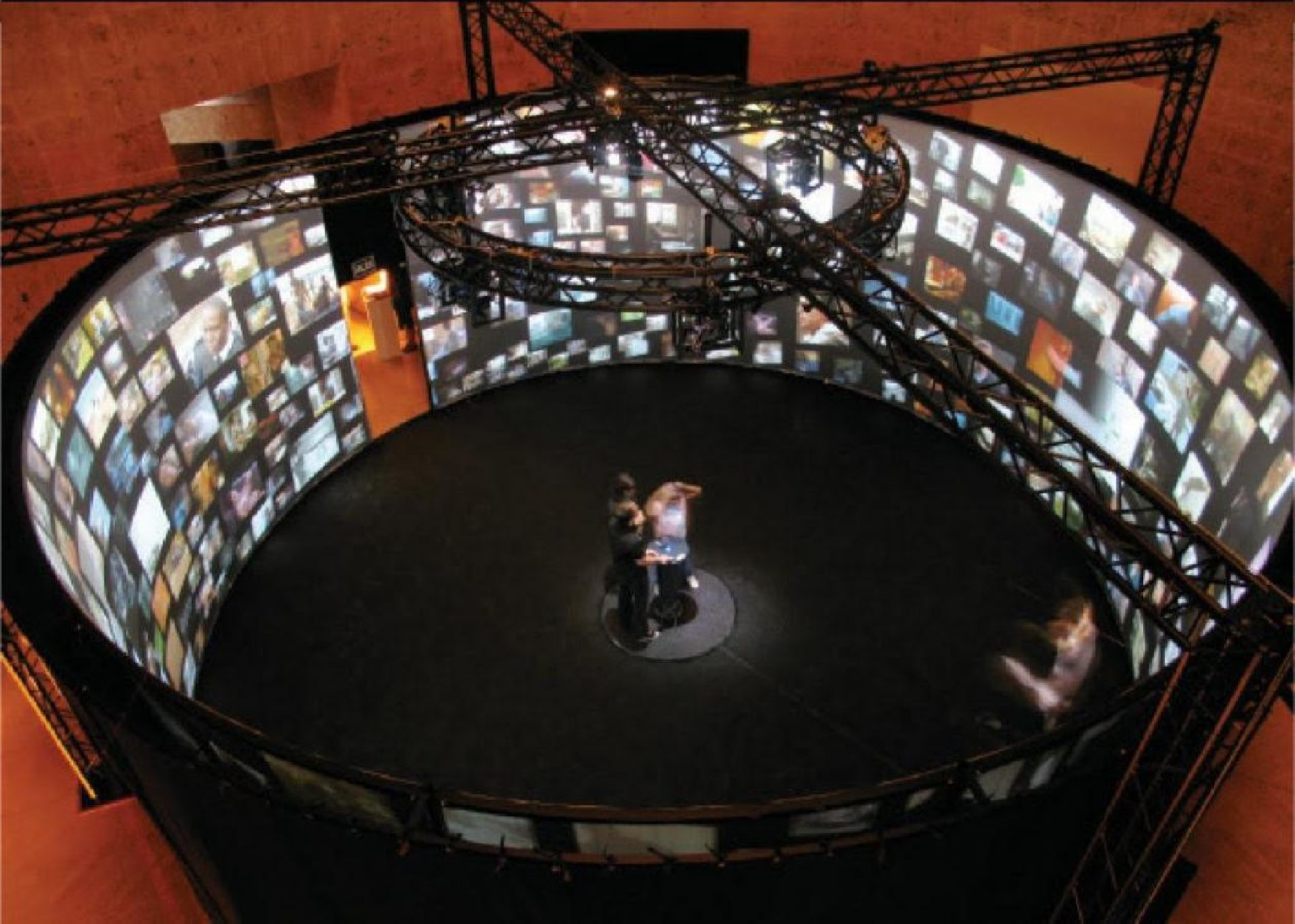}
 \caption{View of the \emph{T\_Visionarium II} installation (credits: Sarah Kenderdine)}
 \label{fig:tvisionarium}
\end{figure}

\emph{T\_Visionarium II} is part of the \emph{T\_Visionarium} project, developed at the UNSW's iCinema Research Centre between 2004 and 2017, resulting in three iterations of the work\footnote{See \url{http://www.icinema.unsw.edu.au/projects/t_visionarium/project-overview/} for a project overview.}. The first version was developed for an inflatable dome structure \parencite{tvisionarium_userguide}, while the more advanced second and third versions use the AVIE system, as it can be seen in Figure \ref{fig:tvisionarium}. 
The interactive and immersive installation explores 24 hours of television footage, segmented, manually annotated (based on a thesaurus) and transformed in a database of more than 20,000 clips of a few seconds each. The system is meant for a single user who navigates it with a touch tablet and with the larger audience witnessing the emergent narrative. Hundreds of clips are simultaneously playing on the 360-degree screen, and when the user-agent selects one, the digital landscape rearranges itself, mapping semantically similar clips closer to the selected item (based on the annotated metadata). The selected clips can be recombined together, rewriting the linear narrative of the original footage (already broken down by the initial segmentation) and resulting in a ``recombinatory" or ``transcriptive" narrative \parencite{brown_performing_2011}.

\begin{figure}[!h]
 \centering
 \includegraphics[width = 0.7\textwidth]{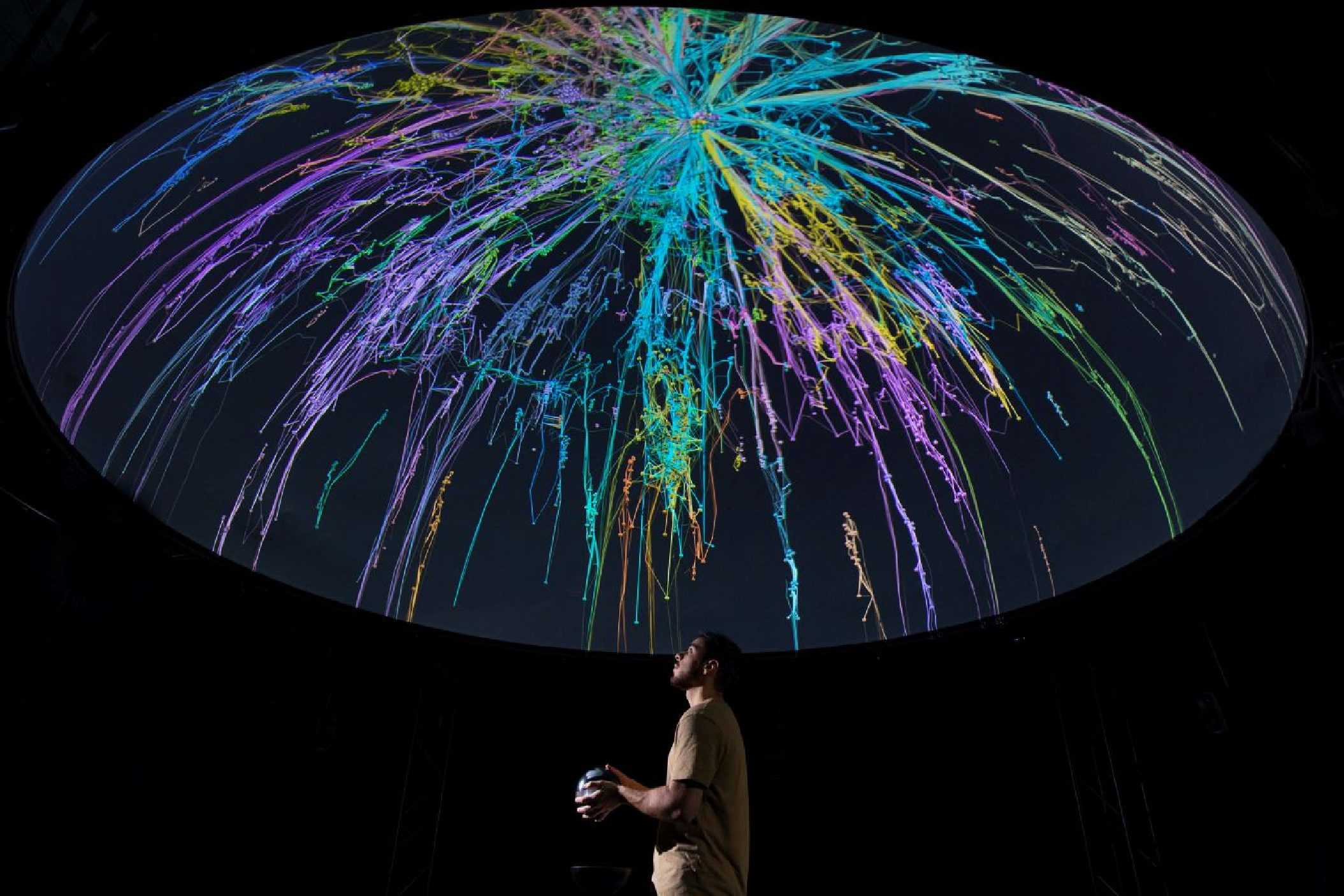}
 \caption{View of the \emph{Jazz Luminaries} installation (credits: Sarah Kenderdine)}
 \label{fig:jazz_luminaries}
\end{figure}

\emph{Jazz Luminaries}, on the other hand, is a much more recent project. Part of the \emph{Infinity Room II} exhibition, the application was developed by EPFL's Laboratory of Experimental Museology in 2019, in a full dome structure of six meters diameter. 13,000 videos of the Montreux Jazz Archive are arranged in a network where nodes represent artists and links their collaborations during the festival. The network can be navigated with a spherical controller, mimicking the structure of the dome, and when passing over a certain node, its corresponding sound excerpts rapidly plays, resulting in an acoustic search akin to radio channel surfing. Lying down under the dome, users thus explore the archive and when an artist is selected, they can choose a specific performance to finally reach a fractal view of the corresponding video (allowing users to appreciate it in spite of their relative position under the dome). As in \emph{T\_Visionarium II}, one spectator drives the experience with the spherical controller, as shown in Figure \ref{fig:jazz_luminaries}, while the others appreciate the unfolding of narrative, reclined under the dome \parencite{kenderdine2021experimental}.

Both installations evidently pivot on a browsing experience browsing experience, where the user-agent explores hundreds of videos simultaneously, gradually revealing details of the collection. The omnidirectional immersion offered by the AVIE system relates to Johnson's idea of the ecological body, as users are ``inside" the dataset, affected and affecting it through their interactions. Similarly, the dome structure recalls a rich history of planetarium structures with the obvious metaphor of the sky vault, a history rich of hundreds of years where the goal is to cover the human field of view. By completely covering the surface of the screen with videos or images, users are naturally inclined to turn their head and look around, as if they were physically exploring a real room filled with archival content, echoing the ecological and phenomenological body through this kinaesthetic paradigm. The first iteration of the \emph{T\_Visionarium} project put even a stronger emphasis on this phenomenological aspect, since the visitor wore a head-tracking system enabling the projected portion of the dome surface, calibrated to the orientation of their head. While turning around, they would therefore visually unveil the extent of the database \parencite{tvisionarium_userguide}. The phenomenon of wandering around in the archive, without a clear path to follow, calls up the model of the ``information flaneur" \parencite{dork2011information}. Curious, creative and critical, the goal of this information-seeker modeled on the 1840s Paris urban flaneur is not to find something in particular but rather to appreciate the collection as a whole, and to be surprised by what they have stumbled upon and ultimately simply be immersed in the archive. Furthermore, this serendipitous search paradigm relates to the idea that ``information is organic" \parencite{serendipity_web_museums}, and hence should be explored in an organic way. 

In \emph{Jazz Luminaries}, such a natural aspect of the data is also suggested through the acoustic search, that draws on the biological body and references the concept of ``tangialities", that are not just related to the touch but include all the five senses \parencite{milekic2002towards}. The social interactions that might arise from a single user driving the experience for a larger public recall the social body as well as Mul's ``third-person's perspective", since one could argue it is the public that is experiencing the full performance, defined by the emergent narrative that the user-agent is creating. There is, as such, no predefined narrative but an infinite combination of items sequences only bounded by viewers' interpretative power. While browsing the archive, it is as if users were ``sculpting" its contents \parencite{kenderdine_cultural_2013}, shaping the experience the way they wish, while being guided by the relationships between elements based on the metadata without being constrained by them. It is clear that such a paradigm could be augmented with modern computational approaches, allowing to generate much more intricacy and thus possibilities in the database, in particular if relying also on more visual features, such as in the SEMIA project illustrated previously \parencite{masson-2020}. Finally, the cultural body is inherent in the exploration of a cultural collection, and entails that prior individual knowledge as well as different cultures are yet another variable in the process of generating narrative. Suffice to say, that a visitor actually having attended one of the Montreux Jazz Festival concerts and then re-experiencing it reclined under the dome will have a much different experience than those discovering the Festival for the first time.

Recalling Aylett and Louchart's narrative theory, \emph{T\_Visionarium II} and \emph{Jazz Luminaries} have a strong contingency in time and space: first, because the narrative is completely generative and determined by users' interactions; secondly due to the important feeling of ``being there" due to the full immersion in the archive and the idea of ``sculpting" it; third, thanks to the high level of interactivity (at least for the user-agent driving the experience). In this way, IVS adopts a form of narrative based on the concept of creating a story through the millions of paths embedded in the latent structures of the collections. Through interactivity and the freedom to explore the horizontal axes of the archive, visitors obtain a strong authorial power on the narrative. In Geert Mul's words, visitors are ``highly productive, in that the appearance of the works changes based on their input" \parencite{mul-2018}, and this entails the clear benefits in the learning experience maintained by constructionist theories. At the same time, it is important to ensure intuitive interaction frameworks are enabled to minimize the learning curve to adopt the system as well as not relying on users having clear goals in mind, something that cannot reliably be expected from casual visitors in a museum setting. In addition, as previously stated, museums are social venues, and interactions between visitors are welcomed and encouraged, such as viewers passing around the spherical controller in \emph{Jazz Luminaries}. Johson's concept of the social body is one of the reasons why multi-user shared spaces in IEs are arguably more interesting than traditional Head-Mounted Displays (HMDs) for IVS in museums spaces. Indeed, even though from a technical point of view, immersion might be higher while using HMDs, the stronger grounding in the physical reality offered by large display screen systems such as the Panorama+ / AVIE allows for a more humanistic approach to the collection, based on social proximity with other visitors. 

The two installations discussed, \emph{T\_Visionarium II} (as well as its predecessor) and \emph{Jazz Luminaries}, have provided clear use cases to illustrate the innovative framework proposed. These projects have their limitations, however, often because they are based on traditional metadata rather than more intrinsic relationships related to strictly visual features (such as colors, visual complexity, movement...) that modern computer vision approaches can unveil. These restrictions are amongst the issues that the next generation of audiovisual archives browsers and projects like \emph{Narratives from the Long Tail: Transforming Access to Audiovisual Archives} are endeavouring to overcome.

\section{Conclusion}

In this article, we present an innovative framework to explore large audiovisual collections through embodied experiences in immersive environments, drawing on Immersive Visual Storytelling theory, Whitelaw's concept of generous interfaces, and Johnson's theory of the body. The need for such a framework is highlighted by archival and humanities scholars as well as the GLAM sector. Two use cases have been analyzed in-depth and critiqued in light of our proposition, supporting our claim that a move towards a more immersive generosity will enhance the experience of visitors engaging with large cultural collections in museum settings through embodied understanding.

The physicality of moving through the collection within which users are immersed, and modifying the way the system presents itself after each interaction, entails the metaphor of sculpting the data, as if it were a raw block of material offered to the viewer to create their own narrative. Furthermore, the social relationships that arise in such shared immersive spaces are key to obtaining a full picture of the embodied understanding of the kinds of cultural
collections that people can experience in a museum. Nonetheless, work remains to be done in the field to improve access to further democratize these collections, defining the research that the \emph{Narratives from the Long Tail: Transforming Access to Audiovisual Archives} project intends to undertake as it maps such possibilities.

\section{Acknowledgements}
This research is supported by the Swiss National Science Foundation through a Sinergia grant for the interdisciplinary project \emph{Narratives from the Long Tail: Transforming Access to Audiovisual Archives}, lead by co-author Prof. Sarah Kenderdine (grant number CRSII5\_198632, see \url{https://www.futurecinema.live/project/} for a project description).

\newpage
\printbibliography

\end{document}